# Dipole spin polarizabilities and gyrations of spin-1 particles in the Duffin-Kemmer-Petiau formalism

N.V. Maksimenko[1], E.V. Vakulina[2], S.M. Kuchin[2]

[1]*The F. Skorina Gomel State University, Gomel, Belarus*

[2]*Affiliated branch of Bryansk State University n.a. Academy Fellow I.G. Petrovskiy, Novozybkov, Russia*

In this paper relativistic-invariant phenomenological Lagrangians of interaction between spin-1 particles and electromagnetic field were obtained in the Duffin-Kemmer-Petiau formalism on the basis of the covariant model that takes into account both spin polarizabilities and gyrations of the above-mentioned particles. It was shown that in the suggested covariant model with regard to the crossing symmetry, spatial parity and gauge invariance conservation laws, definite spin polarizabilities and gyrations of spin-1 particles contribute to the expansion of Compton scattering amplitude, starting from the corresponding orders on energy of pfotons that is in the agreement with low-energy theorems for that process.

PACS numbers: 11.10.Ef.[1]

## Introduction

With the development of the Standard Model of electroweak interaction, new electromagnetic properties of hadrons have been introduced recently. These properties, by analogy with gyration [1,2], are connected with parity violation [3,4]. In their turn, such electromagnetic characteristics as polarizabilities and gyrations are directly related to the inner structure of hadrons and the mechanism of electroweak photon-hadron interactions.

For more reliable determination of polarizabilities and hadron characteristics connected with parity violation, a wide class of electrodynamic processes is used. These processes include real and virtual photons scattering, as wells as two-photon production in hadron-hadron interactions. In this context, the task of consistent relativistic-invariant determination of the contributions of polarizabilities and electroweak characteristics of particles to the electrodynamic processes' amplitudes and cross-sections is of great relevance.

The solution for this task can be found in the framework of relativistic theoretical and field approach to the description of interaction between electromagnetic field and hadrons with the account for polarizabilities (both electromagnetic and electroweak) of the latter. In papers [1, 5-9] covariant techniques describing the interaction between electromagnetic field and hadrons were presented. In such techniques the electromagnetic characteristics of particles are fundamental.

---

[2]elvakulina@yandex.ru

Effective covariant Lagrangian of interaction between electromagnetic field and spin-1/2 particles that takes into account the polarizabilities of the latter was introduced in [1, 10] and has been recently used for fitting the photon-proton scattering experimental data at the energies close to resonance production Δ(1232) [11]. Characterization of electrodynamic processes on the basis of relativistic theoretical and field approaches, which are focused on the obtaining of phenomenological Lagrangians, equations that describe interaction of electromagnetic field with hadrons, as well as the calculation of electrodynamic processes' amplitudes consistent with the Standard Model's low-energy theorems is one of the most effective methods of interaction processes investigation.

Currently there is a number of theoretical papers (see [12-16]) devoted to introduction and calculation of spin polarizabilities of spin-1/2 hadrons that contribute to the series expansion of Compton scattering amplitude at the energies of photons in the third expansion order.

Along with the investigations of spin-1/2 hadrons' polarizabilities, a number of papers present the results of determination and estimation of spin-1 particles' polarizabilities [17-20]. Such particles are characterized by both dipole, spin and tensor polarizabilities.

Low-energy theorems play an important role in the understanding of interaction between electromagnetic field and hadrons. It is stipulated by the fact that they are based on the general concepts of quantum field theory and series expansion of Compton scattering amplitude in powers of photons energy. Currently, one of the most efficient methods of electrodynamic processes investigation is the technique that uses phenomenological Lagrangians obtained in the framework of theoretical and field approaches and consistent with the low-energy theorems that are specified by the Standard Model of electroweak interactions. Construction of such Lagrangians allows to obtain physical interpretation of electromagnetic and electroweak characteristics of hadrons.

In paper [19] low-energy theorems for Compton scattering on a spin-1 particle were obtained. On the basis of these and with the use of techniques for determination of the contribution of spin-1/2 particles' polarizabilities to the amplitudes of electrodynamic processes, one can obtain relativistic-invariant effective Lagrangians and covariant spin structures of two-photon interaction amplitudes with consideration of polarizabilities and electroweak properties

(gyrations) on spin-1 particles. The present paper is entirely devoted to the above-mentioned task.

In paper [21] the construction of the effective relativistic-invariant Lagrangian of interaction between electromagnetic field and particles with constant electric and magnetic dipole moments was performed with the help of dipole moments' anti-symmetric tensor that is independent of electromagnetic field tensor $F_{\mu\nu}$.

The present article uses quantum-field relativistic-invariant Lagrangian, in which a tensor of induced dipole moments is introduced. It means that, in contrast to paper [21], this tensor depends on $F_{\mu\nu}$ [22]. In its turn, polarizabilities tensor [23, 24] is introduced to determine contributions of polarizabilities and gyrations to the low-energy Compton scattering amplitude with provision for particles, spin degrees of freedom. Moreover, we take into account hermiticity requirements, algebra of spin operators and the behavior of tensor components under space and time inversion.

Such phenomenological approach allows to determine the effective relativistic-covariant Lagrangian using the relativistic field consideration of the properties of C-, P- and T-transformations, as wells as the crossing symmetry. It also provides for conformance with the low-energy theorems for Compton scattering on spin-1 particles.

In the present paper the Lagrangian and the amplitude of Compton scattering on the spin-1 particles in the Duffin-Kemmer-Petiau formalism with consideration of their polarizabilities and gyrations were obtained in the framework of covariant theoretical and field approach. The technique presented in papers [5, 22, 25, 26] was used.

## Determination of the spin structure of low-energy amplitude for spin-1 particle Compton scattering

We will follow the paper [27] in order to determine the contributions of polarizabilities and gyrations to the low-energy amplitude of electromagnetic field scattering on spin-1 particle. However, to calculate induced electric $\vec{d}$ and magnetic $\vec{m}$ moments in terms of the electric $\vec{E}$ and magnetic $\vec{H}$ vectors of electromagnetic field strength, we use the following formulas [2]:

$$\vec{d} = 4\pi\hat{\alpha}\vec{E}, \qquad (1)$$
$$\vec{m} = 4\pi\hat{\beta}\vec{H}, \qquad (2)$$

where $\hat{\alpha}$ and $\hat{\beta}$ – are matrices, matrix-elements of which are the tensors of electric and magnetic polarizabilities. Diagonal elements of these matrices are expressed through scalar electric and magnetic polarizabilities:

$$\alpha_{ij} = \alpha_1 \delta_{ij},$$
$$\beta_{ij} = \beta_1 \delta_{ij}.$$

Low-energy amplitude of electromagnetic field scattering that was obtained using formulas (1) and (2) can be presented in the following way [26]:

$$M(\vec{n}_2) = 4\pi\omega^2\{(\vec{e}^{(\lambda_2)*}\hat{\alpha}\vec{e}^{(\lambda_1)}) + (\vec{n}_2\vec{e}^{(\lambda_1)})(\vec{n}_1\hat{\beta}\vec{e}^{(\lambda_2)*}) +$$

$$+(\vec{n}_1\vec{e}^{(\lambda_2)*})(\vec{e}^{(\lambda_1)}\hat{\beta}\vec{n}_2) - (\vec{e}^{(\lambda_2)*}\vec{e}^{(\lambda_1)})(\vec{n}_1\hat{\beta}\vec{n}_2) - (\vec{n}_1\vec{n}_2)(\vec{e}^{(\lambda_1)}\hat{\beta}\vec{e}^{(\lambda_2)*}) +$$

$$+[(\vec{n}_2\vec{n}_1)(\vec{e}^{(\lambda_2)*}\vec{e}^{(\lambda_1)}) - (\vec{n}_2\vec{e}^{(\lambda_1)})(\vec{n}_1\vec{e}^{(\lambda_2)*})]Sp(\hat{\beta})\} \quad (3)$$

Expression (3) includes the following designations: $\omega$ is the incident wave frequency, $\vec{n}_1 = \frac{\vec{k}_1}{|\vec{k}_1|}$, $\vec{e}^{(\lambda_1)}$ and $\vec{k}_1$ are correspondingly the polarization and wave vectors of the incident wave.

According to the definitions of $\vec{d}$ and $\vec{m}$ presented in (1) and (2), it follows that $\hat{\alpha}$ and $\hat{\beta}$ should satisfy the hermiticity requirement. Taking into account this requirement as well as the algebra of spin-1 operators $\hat{S}_i$ [19] we can obtain the following:

$$[\hat{S}_i, \hat{S}_j] = i\delta_{ijk}\hat{S}_k, \quad (4)$$

$$\hat{S}_i\hat{S}_j\hat{S}_k = i\delta_{ijk} + \frac{1}{2}(\hat{S}_i\delta_{jk} + \hat{S}_k\delta_{ij}) + \frac{i}{2}\delta_{ikl}(\hat{S}_j\hat{S}_l + \hat{S}_l\hat{S}_j), \quad (5)$$

$\hat{\alpha}$ and $\hat{\beta}$ operators can be presented in the following way [26]:

$$\alpha_{ij} = \alpha_1\delta_{ij} + i\alpha_2\delta_{ijk}\hat{S}_k + i\mathcal{X}_E\delta_{ijk}\partial_k + \bar{\bar{\alpha}}(\hat{S}_i\hat{S}_j + \hat{S}_j\hat{S}_i), \quad (6)$$

$$\beta_{ij} = \beta_1\delta_{ij} + i\beta_2\delta_{ijk}\hat{S}_k + i\mathcal{X}_M\delta_{ijk}\partial_k + \bar{\bar{\beta}}(\hat{S}_i\hat{S}_j + \hat{S}_j\hat{S}_i), \quad (7)$$

where $i, j, k$ and $l$ can take the value of 1,2 or 3, while $\delta_{ijk}$- is the three-dimensional Levi-Civita tensor.

In formulas (6) and (7) $\alpha_1$ and $\beta_1$ – are scalar dipole electric and magnetic polarizabilities correspondingly, $\bar{\bar{\alpha}}$ and $\bar{\bar{\beta}}$ - are tensor polarizabilities, $\alpha_2$ and $\beta_2$– are spin dipole polarizabilities, while $\mathcal{X}_E$ and $\mathcal{X}_M$ - are correspondingly electric and magnetic gyrations. As a consequence of crossing symmetry, $\alpha_2$, $\beta_2$ and $\mathcal{X}_E$, $\mathcal{X}_M$ have non-zero contribution to the amplitude of Compton scattering in the third expansion order of the photons energy.

As it was shown in [26], by substituting formulas (6) and (7) into (3) and taking into account the contributions of $\alpha, \beta, \bar{\alpha}$ and $\bar{\bar{\beta}}$ polarizabilities, one can obtain the scattering amplitude in the second expansion order of the photons energy. It coincides with beyond the Born part of the amplitude and is due to the low-energy theorem [19].

Let's determine relativistic-invariant spin structures of the effective Lagrangian and the amplitudes of Compton scattering on spin-1 particles with the help of covariant representation of (6) and (7) in the Duffin-Kemmer-Petiau (DKP) formalism following paper [26].

The DKP equations for an unbounded spin-1 particle have the following form [28]:

$$(\beta_\mu \vec{\partial}_\mu + m)\psi(x) = 0, \qquad (8)$$

$$\bar{\psi}(x)(\beta_\mu \overleftarrow{\partial}_\mu - m) = 0, \qquad (9)$$

where $\psi(x)$ and $\bar{\psi}(x) = \psi^+(x)\eta$ - are ten-dimensional functions of particles, $\eta = 2\left(\beta_4^{(10)}\right)^2 - I$, vectors over derivatives $\partial_\mu$ show the direction of their action, while four-dimensional vector is defined as $a_\mu\{\vec{a}, ia_0\}$. In formulas (8) and (9) $\beta_\mu$ - are ten-dimensional DKP matrices that satisfy the following commutation rules:

$$\beta_\mu \beta_\nu \beta_\rho + \beta_\rho \beta_\nu \beta_\mu = \delta_{\mu\nu}\beta_\rho + \delta_{\rho\nu}\beta_\mu.$$

In the framework of theoretical and field covariant approach the effective Lagrangian of interaction between electromagnetic field and spin-1 particle with provision for polarizabilities has the form [5, 8, 26]:

$$L = -\frac{\pi}{2m}\bar{\psi}[\beta_\nu \hat{L}_{\nu\sigma}\overleftrightarrow{\partial}_\sigma + \hat{L}_{\nu\sigma}\beta_\nu \overleftrightarrow{\partial}_\sigma]\psi, \qquad (10)$$

where $\overleftrightarrow{\partial}_\sigma = \vec{\partial}_\sigma - \overleftarrow{\partial}_\sigma$.

The formula (10) for the Lagrangian includes tensor $\hat{L}_{\nu\sigma}$, which is expressed in terms of polarizabilities and gyrations as:

$$\hat{L}_{\nu\sigma}(\alpha, \chi_E) = \hat{L}_{\nu\sigma}(\alpha_1) + \hat{L}_{\nu\sigma}(\bar{\alpha}) + \hat{L}_{\nu\sigma}(\alpha_2) + \hat{L}_{\nu\sigma}(\chi_E), \qquad (11)$$

$$\hat{L}_{\nu\sigma}(\alpha, \chi_M) = \hat{L}_{\nu\sigma}(\beta_1) + \hat{L}_{\nu\sigma}(\bar{\bar{\beta}}) + \hat{L}_{\nu\sigma}(\beta_2) + \hat{L}_{\nu\sigma}(\chi_M), \qquad (12)$$

In order to determine the influence of crossing symmetry on the contributions of spin polarizabilities and gyrations to the Compton scattering amplitude in dipole representation we will transform tensors (11) as (see [22]):

$$\hat{L}_{\nu\sigma}(\alpha_1) + \hat{L}_{\nu\sigma}(\bar{\bar{\alpha}}) = F_{\nu\mu}\hat{\alpha}^{\mu\rho}(\alpha_1)F_{\rho\sigma} + F_{\nu\mu}\hat{\alpha}^{\mu\rho}(\bar{\bar{\alpha}})F_{\rho\sigma} \qquad (13)$$

$$\hat{L}_{\nu\sigma}(\alpha_2) + \hat{L}_{\nu\sigma}(\chi_E) = F_{\nu\mu}\overleftrightarrow{\partial}_\lambda F_{\rho\sigma}\hat{k}_{\mu\rho\lambda}(\alpha_2) + F_{\nu\mu}\overleftrightarrow{\partial}_\lambda F_{\rho\sigma}\hat{k}_{\mu\rho\lambda}(\chi_E) \qquad (14)$$

Derivatives in equation (14) operate only on the tensors of electromagnetic field

$$F_{\mu\nu} = \partial_\mu A_\nu - \partial_\nu A_\mu,$$

Tensors $\hat{\alpha}^{\mu\rho}(\alpha_1)$ and $\hat{\alpha}^{\mu\rho}(\bar{\bar{\alpha}})$, as well as $\hat{k}_{\mu\rho\lambda}(\alpha_2)$ and $\hat{k}_{\mu\rho\lambda}(\chi_E)$ are the covariant generalization of tensors that appear in the right part of formula (6). They have the following form:

$$\hat{\alpha}_{\mu\rho} = \alpha_1 \delta_{\mu\rho} + \bar{\bar{\alpha}}(\widehat{W}_\mu \widehat{W}_\rho + \widehat{W}_\rho \widehat{W}_\mu), \qquad (15)$$

$$\hat{k}_{\mu\rho\lambda} = \frac{i\alpha_2}{2m}\delta_{\mu\rho\lambda\kappa}\widehat{W}_\kappa + \frac{i\chi_E}{2m}\delta_{\mu\rho\lambda\kappa}\overleftrightarrow{\partial}_\kappa. \qquad (16)$$

In equations (15) and (16) the definition of covariant spin vector is used. This vector can be expressed in terms of $\beta_\nu$ matrices (see [28]):

$$W_\mu = -\frac{i}{4m}\delta_{\mu\kappa\delta\eta}\hat{J}^{[\delta\eta]}\overleftrightarrow{\partial}_\kappa,$$

where $\hat{J}^{[\delta\eta]} = \beta_\delta\beta_\eta - \beta_\eta\beta_\delta$. All derivatives found in (15) and (16) operate on wave functions $\psi$ and $\bar{\psi}$.

Tensor (12) is defined in a similar way. One just needs to introduce constants $\beta_1, \beta_2, \bar{\bar{\beta}}$ and $\chi_M$, in formulas (13)-(14) and make a replacement

$$F_{\nu\mu} \to \tilde{F}_{\nu\mu},$$

where

$$\tilde{F}_{\mu\nu} = \frac{i}{2}\delta_{\mu\nu\rho\sigma}F_{\rho\sigma}.$$

Let's now determine the spin structures of the amplitude of Compton scattering on spin-1 particle with provision for polarizabilities and gyrations. We will take Lagrangian (10) as a basis and follow the procedure presented in paper [28]:

$$\langle k_2, p_2|\hat{S}|k_1, p_1\rangle = \frac{im\delta(k_1+p_1-k_2-p_2)}{(2\pi)^2\sqrt{4\omega_1\omega_2 E_1 E_2}}\mathrm{M}, \qquad (17)$$

here M is the Compton scattering amplitude that represents the sum of polarizabilities and gyrations contributions according to formulas (11) and (12).

As it was shown in [26], the contribution of $\alpha$, $\beta$ and $\bar{\bar{\alpha}}$, $\bar{\bar{\beta}}$ is expressed as a sum of amplitudes

$$M_1 = M_1(\alpha,\beta) + M_1(\bar{\bar{\alpha}},\bar{\bar{\beta}}). \qquad (18)$$

Spin structure $M(\alpha, \beta)$ in equation (18) has the following form:

$$M_1(\alpha, \beta) = \left(-\frac{2\pi i}{m}\right)\{\alpha\left[F_{\nu\mu}^{(2)}F_{\mu\sigma}^{(1)} + F_{\nu\mu}^{(1)}F_{\mu\sigma}^{(2)}\right] +$$

$$+\beta\left[\tilde{F}_{\nu\mu}^{(2)}\tilde{F}_{\mu\sigma}^{(1)} + \tilde{F}_{\nu\mu}^{(1)}\tilde{F}_{\mu\sigma}^{(2)}\right]\}P_\sigma\bar{\psi}^{(r_2)}(p_2)\beta_\nu\psi^{(r_1)}(p_1). \qquad (19)$$

In its turn, structure $M(\bar{\bar{\alpha}}, \bar{\bar{\beta}})$ is determined as:

$$M_1(\bar{\bar{\alpha}}, \bar{\bar{\beta}}) = \left(-\frac{\pi i}{m}\right)\{\bar{\bar{\alpha}}\left[F_{\nu\mu}^{(2)}F_{\mu\sigma}^{(1)} + F_{\nu\mu}^{(1)}F_{\mu\sigma}^{(2)}\right] +$$

$$+\bar{\bar{\beta}}\left[\tilde{F}_{\nu\mu}^{(2)}\tilde{F}_{\mu\sigma}^{(1)} + \tilde{F}_{\nu\mu}^{(1)}\tilde{F}_{\mu\sigma}^{(2)}\right]\}P_\sigma\bar{\psi}^{(r_2)}(p_2)[\beta_\nu\{\widehat{W}_\mu, \widehat{W}_\rho\} + \{\widehat{W}_\mu, \widehat{W}_\rho\}\beta_\nu]\psi^{(r_1)}(p_1). (20)$$

Equations (19) and (20) include the following designations:

$$F_{\nu\mu}^{(2)} = k_{2\nu}e_\mu^{(\lambda_2)*} - k_{2\mu}e_\nu^{(\lambda_2)*},$$

$$F_{\mu\sigma}^{(1)} = k_{1\mu}e_\sigma^{(\lambda_1)} - k_{1\sigma}e_\mu^{(\lambda_1)},$$

where $\tilde{F}_{\nu\mu}^{(2)} = \frac{i}{2}\delta_{\nu\mu\varkappa\delta}F_{\varkappa\delta}^{(2)}$, $P_\sigma = \frac{1}{2}(p_1 + p_2)_\sigma$, $p_1$ and $p_2$ - are the momenta of initial and final spin-1 particles correspondingly.

Ten-dimensional wave functions in the DKP formalism are introduced using complete matrix algebra elements $\varepsilon^{AB}$ [28]

$$\psi^{(r)}(p) = \psi_\mu^{(r)}(p)\varepsilon^{\mu 1} + \frac{1}{2}\psi_{[\mu\nu]}^{(r)}(p)\varepsilon^{[\mu\nu]1}.$$

In this formula

$$\psi_\mu^{(r)}(p) = \frac{i}{\sqrt{2}}\lambda_\mu^{(r)},$$

$$\psi_{[\mu\nu]}^{(r)}(p) = -\frac{1}{\sqrt{2}m}\left(p_\mu\lambda_\nu^{(r)} - \lambda_\mu^{(r)}p_\nu\right),$$

$\lambda_\mu^{(r)}$ – are the components of polarization vectors of spin-1 particle, while $\varepsilon^{AB}$ – are the elements of complete matrix algebra [28]:

$$(\varepsilon^{AB})_{CD} = \delta_{AC}\delta_{BD}, \varepsilon^{AB}\varepsilon^{CD} = \delta_{BC}\varepsilon^{AD},$$

where for spin-1 particle indices $A, B, C, D = \mu, [\rho\sigma]$, while square brackets stand for the anti-symmetry with respect to indices $\rho$ and $\sigma$.

Wave functions $\bar{\psi}^{(r)}(p)$ that are conjugate with respect to $\psi^{(r)}(p)$ are expressed in the following way (taking into account $\eta$ matrix):

$$\bar{\psi}^{(r)}(p) = \psi^+(p)\eta = \left(-\frac{i}{\sqrt{2}}\right)\left[\dot{\lambda}_\mu^{(r)}\varepsilon^{1\mu} + \frac{i}{2m}\varepsilon^{1[\mu\nu]}\left(p_\mu\dot{\lambda}_\nu^{(r)} - p_\nu\dot{\lambda}_\mu^{(r)}\right)\right],$$

where $\dot{\lambda}_\mu^{(r)}\left\{\lambda_i^{(r)*}, \lambda_4^{(r)}\right\}$.

Let's now determine the spin structures of the amplitudes with provision for the contributions of dipole spin polarizabilities $\alpha_2, \beta_2$ and gyrations $\chi_E, \chi_M$, i.e.

$$M_2 = M_2(\alpha_2, \beta_2) + M_2(\chi_E, \chi_M).$$

Using the summands $\hat{L}_{\nu\sigma}(\alpha_2), \hat{L}_{\nu\sigma}(\chi_E), \hat{L}_{\nu\sigma}(\beta_2)$ (11) and $\hat{L}_{\nu\sigma}(\chi_M)$ (12), of the Lagrangian, as well as the previous technique for determination of polarizations contributions to the Compton scattering amplitude, one can find:

$$M_2(\alpha_2, \beta_2) = \frac{\pi}{m}(k_1 + k_2)_\lambda \delta_{\mu\rho\lambda\kappa}\{\alpha_2\left[F_{\nu\mu}^{(2)}F_{\rho\sigma}^{(1)} - F_{\nu\mu}^{(1)}F_{\rho\sigma}^{(2)}\right] +$$

$$+\beta_2\left[\tilde{F}_{\nu\mu}^{(2)}\tilde{F}_{\rho\sigma}^{(1)} - \tilde{F}_{\nu\mu}^{(1)}\tilde{F}_{\rho\sigma}^{(2)}\right]\}\bar{\psi}^{(r_2)}(p_2)[\beta_\nu\widehat{W}_k + \widehat{W}_k\beta_\nu]P_\sigma\psi^{(r_1)}(p_1). \quad (21)$$

Amplitude (21) in the target's rest frame and with the neglect of the target particle's recoil can be expressed as:

$$M_2(\alpha_2, \beta_2) = 4i\pi(\omega_1 + \omega_2)\omega_1\omega_2\vec{\lambda}^{(r_2)*} \cdot$$

$$\cdot\{\alpha_2(\vec{S}[\vec{e}^{(\lambda_2)*}\vec{e}^{(\lambda_1)}]) + \beta_2(\vec{S}[\vec{n}_2\vec{e}^{(\lambda_2)*}][\vec{n}_1\vec{e}^{(\lambda_1)}])\}\vec{\lambda}^{(r_1)} \quad (22)$$

Formulas (21) and (22) imply that dipole spin polarizabilities $\alpha_2$ and $\beta_2$ contribute to the amplitude of Compton scattering on spin-1 particle in the third expansion order (series expansion in the energy of photons), while the crossing symmetry requirements and parity conservation (with respect to space inversion) rules are satisfied.

Using the above-introduced technique for constructing covariant blocks of the effective Lagrangian with provision for the crossing symmetry and parity violation, we can obtain the second summand of the amplitude that depends on the contributions of electric and magnetic gyrations:

$$M_2(\chi_E, \chi_M) = \frac{2\pi i}{m^2}(k_1 + k_2)_\lambda \delta_{\mu\rho\lambda\kappa}\{\chi_E\left[F_{\nu\mu}^{(2)}F_{\rho\sigma}^{(1)} - F_{\nu\mu}^{(1)}F_{\rho\sigma}^{(2)}\right] +$$

$$+\chi_M\left[\tilde{F}_{\nu\mu}^{(2)}\tilde{F}_{\rho\sigma}^{(1)} - \tilde{F}_{\nu\mu}^{(1)}\tilde{F}_{\rho\sigma}^{(2)}\right]\}P_k P_\sigma\bar{\psi}^{(r_2)}(p_2)\beta_\nu\psi^{(r_1)}(p_1). \quad (23)$$

If we use approximation $\vec{P} = 0$, in equation (23), i.e. we consider the particle to be at rest and neglect its recoil momentum, the formula (23) can be rewritten in the following way:

$$M_2(\chi_E, \chi_M) = 4\pi\omega_1\omega_2\big(\vec{\lambda}^{(r_2)*}\vec{\lambda}^{(r_1)}\big)\{\chi_E(\vec{k}_1 + \vec{k}_2)\big[\vec{e}^{(\lambda_2)*}\vec{e}^{(\lambda_1)}\big] +$$

$$+\chi_M(\vec{k}_1 + \vec{k}_2)\big[\vec{\Sigma}_2\vec{\Sigma}_1\big]\},$$

where $\vec{\Sigma}_2 = \big[\vec{n}_2\vec{e}^{(\lambda_2)*}\big]$, $\vec{\Sigma}_1 = \big[\vec{n}_1\vec{e}^{(\lambda_1)}\big]$.

## Conclusion

Hence, we determined the contributions of polarizabilities to the low-energy Compton scattering amplitude with provision for the spin degrees of freedom of particles by transforming the polarizabilities tensor that satisfies both hermiticity requirement and spin algebra. This tensor is also invariant with respect to space inversion transformations.

The relativistic-covariant form of contributions of spin and tensor polarizabilities, as well as gyrations to the Compton scattering amplitude in the DKP formalism was found.

The effective Lagrangian that takes into account the changes of spin structures during space inversion transformations and considers the crossing symmetry of Compton scattering amplitude on spin-1 particle was obtained in the DKP formalism using theoretical and field relativistic generalization. The coordination of this amplitude with the low-energy theorems was performed as well.

______________________________________